\documentclass[final,times]{elsarticle}

\usepackage[utf8]{inputenc}
\usepackage{graphicx}
\usepackage{subfig}
\usepackage{caption}
\usepackage{booktabs}
\usepackage{xcolor}
\usepackage{numprint}
\usepackage{comment}
\usepackage{algorithm2e}
\usepackage{geometry}
\usepackage{makecell}
\usepackage{url}

\newcommand{\added}[1]{%
   {\color{blue}#1%
   }%
 }
\renewcommand{\added}[1]{#1}
\newcommand{\removed}[1]{%
 {\color{red}#1%
 }%
}
\renewcommand{\removed}[1]{}

\journal{Pervasive and Mobile Computing}

\begin{document}

\begin{frontmatter}
 
\title{Energy consumption of smartphones and IoT devices when using different versions of the HTTP protocol}

\author[dip]{Chiara~Caiazza}
\ead{chiara.caiazza@phd.unipi.it}
\author[iit]{Valerio~Luconi}
\ead{valerio.luconi@iit.cnr.it}
\author[dip]{Alessio~Vecchio\corref{cor1}}
\ead{alessio.vecchio@unipi.it}

\cortext[cor1]{Corresponding author}
\address[dip]{Dip. di Ing. dell'Informazione, Universit\`a di Pisa, Largo L. Lazzarino 1, 56122 Pisa, Italy}
\address[iit]{Istituto di Informatica e Telematica, Consiglio Nazionale delle Ricerche, Via G. Moruzzi, 1, 56124 Pisa, Italy}

\begin{abstract}
HTTP is frequently used by smartphones and IoT devices to access information and Web services. Nowadays, HTTP is used in three major versions, each introducing significant changes with respect to the previous one. We evaluated the energy consumption of the major versions of the HTTP protocol when used in the communication between energy-constrained devices and cloud-based or edge-based services. Experimental results show that in a machine-to-machine communication scenario, for the considered client devices -- a smartphone and a Single Board Computer -- and for a number of cloud/edge services and facilities,  HTTP/3 frequently requires more energy than the previous versions of the protocol. The focus of our analysis is on machine-to-machine communication, but to obtain a broader view we also considered a client-server interaction pattern that is more browsing-like. In this case, HTTP/3 can be more energy efficient than the other versions.

\end{abstract}

\begin{keyword}
energy, HTTP, edge, cloud, measurements, IoT, smartphone.
\end{keyword}

\end{frontmatter}

\section{Introduction}

The Hypertext Transfer Protocol (HTTP) is used not only for accessing websites but also for communicating with a wide range of networked services. The current iteration of the protocol, HTTP/3, has been recently standardized, and it is now being adopted by an always-growing fraction of websites~\cite{http3}. According to the HTTP Archive, HTTP/3 support on requests has grown from 10\% in 2021 to 15\% in 2022 \cite{httparchive}. The adoption has been particularly significant within large Internet companies and Content Delivery Networks (CDNs), such as Cloudflare and Google~\cite{Perna22:first}. 

HTTP/3 introduces significant changes compared to the previous versions of the protocol. In particular, it uses QUIC as the transport protocol whereas the previous versions were based on TCP. The QUIC protocol was originally designed by Google and it is now an IETF standard. QUIC, which is based on UDP, allows faster handshaking and adopts different acknowledgment mechanisms and flow control schemes compared to TCP \cite{46403}. Several studies in recent years evaluated the performance of the different versions of HTTP, for example in terms of page loading speed in production websites~\cite{Perna22:first}, or compared to MQTT when used in the Internet of Things (IoT) domain~\cite{saif2021apure}. 

Nowadays, most of the services accessed by clients are hosted in cloud-based facilities. Recent reports highlighted that the transition to cloud and multicloud infrastructures is accelerating, with 78\% of organizations willing to host more than 40\% of their workloads in the cloud by 2025~\cite{cisconetworkingtrends}. In this context, a great portion of traffic is represented by IoT-produced data~\cite{Nguyenan2021:iot}. Machine-to-machine connections will represent almost half of the total devices and connections by the end of 2023, and a great portion of mobile traffic will be represented by machine-to-machine communication~\cite{ciscointernetreport}. The last years were also characterized by the introduction of a novel network architecture, edge computing, where computational resources are pushed closer to the end-users. The reduced latency between users and edge-based computational resources allows better performances in applications such as AR/VR~\cite{erol2018caching}, video analysis~\cite{8467992}, and IoT-related applications. In fact, IoT is considered by enterprises and telecom operators as one of the top use cases and business drivers for edge computing~\cite{hredge}. Edge computing has been \added{interpreted} in different ways: as a general principle to push processing at the fringes of the network, or as a standardized network architecture like the one proposed by ETSI Multi-access Edge Computing \cite{giust2018mec}. Also, the closeness to the users has been interpreted in different ways, now converging to the idea of an edge-cloud continuum where resources can be located~\cite{Caiazza2022:edge}.

In this paper, we study the differences in terms of energy consumption of the major versions of the HTTP protocol. The study concentrates on energy consumption of smartphones and IoT devices in a cloud/edge-computing scenario, for the following reasons: $i)$ smartphones are nowadays much more popular than personal computers when accessing online services; $ii$) content and services are generally hosted in the cloud or provided by Content Delivery Networks (CDNs); $iii)$ edge computing will play an always more relevant role in the future, as low latency cannot be reached without reducing the length of the path traveled to access information, $iv)$ edge-computing is considered to be an enabling technology for many IoT applications, as it allows to process, aggregate, and filter the data produced by Things as close as possible to the source \cite{7574435}.
The main contributions can be summarized as follows:
\begin{itemize}
    \item We experimentally evaluate different versions of the HTTP protocol focusing on energy consumption. While other performance metrics for different HTTP versions have been studied~\cite{Perna22:first, Yu21:dissecting, Kosar2019:energy}, there are no studies concentrating solely on energy consumption, which we believe can be the most important metric for certain classes of applications, e.g. IoT. For this reason, our primary focus is on machine-to-machine communication. The comparison between the different versions of the HTTP protocols has been carried out starting from the null hypothesis that their consumption is the same. We considered several scenarios, and in a number of them, the null hypothesis is rejected. This, in practice, means that the differences in the energy consumption of the protocol versions are statistically significant.
    
    \item We explored a variety of scenarios, using both commercial edge/cloud servers, and private resources. We considered the main energy-constrained platforms, namely smartphones and Single Board Computers (SBCs), and multiple configurations of operational parameters, such as the client-server distance in the delay space, the payload size, and the degree of parallelism of requests.

    \item The experimental evaluation is extended to a browsing-like communication pattern, which is representative of more traditional usage of the HTTP protocol.
\end{itemize}

Experimental results show that, in a machine-to-machine communication scenario, most of the time HTTP/3 needs more energy compared to the previous versions of the protocol. This is not confirmed in the browser-like scenario, as the execution of multiple requests back-to-back seems to help HTTP/3 be more energy efficient than the other protocol versions.
We believe that our study is extremely relevant, as energy consumption is a crucial factor for both smartphones and IoT devices, as they are all battery-operated. Especially for IoT devices, the energy expenditure can be more important than the other performance metrics, such as throughput, thus we believe that a study considering just energy consumption can help other researchers and practitioners design their communication protocols more wisely.

\section{Related Work and Motivation}

Our work is at the intersection of three domains: $i)$ the energy consumption in the edge-cloud continuum, $ii)$ the impact of communication on energy consumption, and $iii)$ the evaluation of the HTTP protocol versions. 

\subsection{Energy consumption in the edge-cloud continuum}

The specific problem of energy consumption in the edge-cloud continuum has been tackled from various perspectives. Various works have focused on analyzing the energy consumption of data centers~\cite{Katal22:energy}. A significant amount of work explored the possibilities given by computation offloading, i.e. delegating CPU-intensive tasks to edge or cloud servers \cite{TEREFE201675}. Task offloading in an edge/cloud environment for minimizing energy consumption has been evaluated in~\cite{Baliga2007:energy, Rahimi14:mobile, Zahed2020:green,Sarkar2018:assessment, Pei2020:energy, Zhang2018:energy,Hu2016:quantifying,SHAHRYARI2021101395}. Other works focused on estimating the energy consumption of communication on the edge/cloud continuum, from both a theoretical and experimental point of view~\cite{Caiazza2022:edge, Caiazza2022:saving}. In this work, we consider just HTTP-based communication, without computation offloading. 

\subsection{Impact of communication on energy consumption}

The energy consumption of mobile devices when communicating has been widely studied in the past years. Thanks to energy models of the LTE interface, cellular communications were shown to be responsible for a significant fraction of the overall energy consumption in smartphones~\cite{Huang2012:close, Chen2015:smartphone}. This finding stimulated a deeper exploration of the impact of latency on energy consumption. In particular, some works aimed at finding a balance between energy consumption and latency requirements~\cite{Tseng2016:delay, Mehmood2019:power, Brand2020:adaptive}.

Other works have then focused on the energy consumption of Internet protocols at multiple levels of the stack~\cite{Cengiz2015:review}. However, most of such works were aimed at designing energy-aware protocols or promoting an energy-aware mobile application design. In~\cite{Zorzi2001:energy}, the authors analytically modeled the energy consumption of various TCP versions, showing that in wireless scenarios certain TCP versions are more energy-saving than others. An energy consumption profile of the DASH protocol in LTE networks has been provided in~\cite{Zhang2016:profiling} via the use of a power monitor. The authors characterized the energy consumption under various conditions (e.g. video resolution, segment length, buffer size, signal strength), to show that the energy consumption can be reduced with adequately tuned system parameters.

\subsection{Evaluation of the HTTP versions}
When it comes to evaluating different HTTP versions, the vast majority of works focused on \added{classic }performance \added{indices like throughput and delay}~\cite{Barford99:performance, Saif21:early, Marx20:same}. \added{The first studies highlighted how early implementations of QUIC and HTTP/3 showed improved performance with respect to TCP and TLSv1.2, but these improvements were marginal when comparing to TCP and TLSv1.3~\cite{Saif21:early}. In~\cite{Moulay2021:experimental}, QUIC and TCP were evaluated in mobile networks of different European countries. The authors showed that when using no application layer protocol and with a download size of 1~MB, TCP obtained better performance in some countries, whereas in other countries QUIC performed better. However, when comparing application layer protocols such as HTTP/2 and HTTP/3 in a browsing-like scenario with smaller file sizes, HTTP/3 showed better performance. Similar findings are reported in~\cite{Shreedhar2022:evaluating}, where authors showed how earlier versions of HTTP with TCP obtain higher throughput when downloading large files.} In~\cite{Perna22:first}, the authors compared HTTP/1.1, HTTP/2, and HTTP/3 in multiple scenarios \added{using a browsing-like traffic pattern and browsing-related performance metrics}, showing that HTTP/3 can obtain higher performance \added{in some of them}, especially in mobile network conditions. \added{However, they found that in specific configurations, especially with larger websites, HTTP/2 obtained better performance.} \removed{However, these performance improvements could be due to specific optimizations performed by service providers. }Similar findings are reported in~\cite{Yu21:dissecting}, where the authors compare QUIC and TCP, by using HTTP/3 and HTTP/2 as application-layer protocols, respectively. The authors highlight how performance differences can be due to library implementations and service providers' system optimizations. \added{Other works have indeed pointed out that different parameters of operation can lead to substantial differences in the measured throughput and delay of different HTTP versions. For example, different congestion control algorithms of the QUIC protocol can obtain different throughputs and delays~\cite{Haile2022:performance}. Even different implementations of the same algorithm can lead to very different results, as pointed out in~\cite{Mishra2022:understanding}, where the authors compared multiple implementations of popular congestion control algorithms with their kernel counterparts. Similar considerations about congestion control algorithms are reported in~\cite{Moulay2021:experimental}. In general, since QUIC and HTTP/3 are still a work in progress, there is still no standard reference implementation, and different libraries can obtain very different performance results~\cite{Dubec2023:performance}.}

A few works have analyzed HTTP from the standpoint of energy consumption. In~\cite{Li14:empirical}, the authors showed that HTTP requests are a major contributor to the energy consumption of mobile applications, and in subsequent works, the same authors showed how the energy consumption of HTTP requests can be optimized by bundling them~\cite{Li15:optimizing, Li16:automated}. In~\cite{Anwar20:should}, Anwar, et al. compared different HTTP libraries for Android to find that they differ in energy consumption. In this specific case, they did not find a significant correlation between the energy consumption of a library and its classic performance indices. It has to be noted that the authors did not focus on specific HTTP versions.

\subsection{Motivation}

In this paper, we provide an experimental evaluation of different versions of the HTTP protocol from the point of view of energy consumption \added{on the client side}. The study is based on an edge/cloud computing scenario as it is an increasingly adopted architectural solution in current and future generation networks. To the best of our knowledge, this is the first time that such a combined analysis has been performed.

\section{Scenarios and Setup}
\label{sec:setup}

\begin{figure}[!t]
    \centering
    \includegraphics[width=0.4\textwidth]{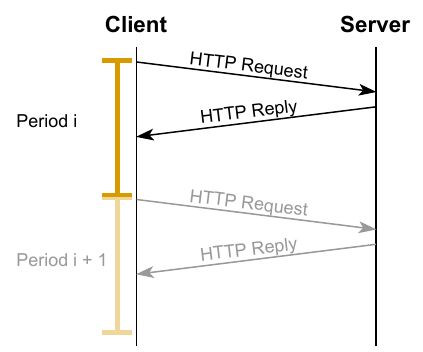}
    \caption{The communication scheme between client and server.}
    \label{fig:client-serverInteraction}
\end{figure}

\begin{table}[!t]
    \centering
    \caption{Devices used in the experiments.}
                \begin{tabular}{lrr}
                    \toprule
                      & Smartphone & Single Board Computer\\
                     \hline
                     Model & Google Nexus 5 &  Raspberry PI 3B+\\
                     OS & Android 6 & Raspberry OS - Buster \\
                     RAM & 2 GB &  1 GB      \\
                     Transceiver & \makecell[r]{Qualcomm \\ Snapdragon™ 800} & \makecell[r]{SIM7600G-H global\\ band 4G module} \\
                     \bottomrule
                \end{tabular}
    \label{tab:devices}
\end{table}

In this study, we evaluate the energy spent by a device to interact with a server in the edge-cloud continuum using the major versions of the HTTP protocol (namely, versions 1.1, 2, and 3). Communication is supposed to take place according to the client-server paradigm. The client makes periodic requests to the server, as illustrated in Figure~\ref{fig:client-serverInteraction}. After the HTTP reply has been entirely received, the client just waits until the start of the next period. In our scenarios, we consider two types of devices, smartphones and SBCs. The first category is relevant as smartphones are the devices that are most frequently used in our everyday life. The second category is representative of IoT clients possibly interacting with a remote server, for example, to upload data or to interact with a service according to a polling approach.

\begin{figure}[!t]
    \centering
    \includegraphics[width=0.6\textwidth]{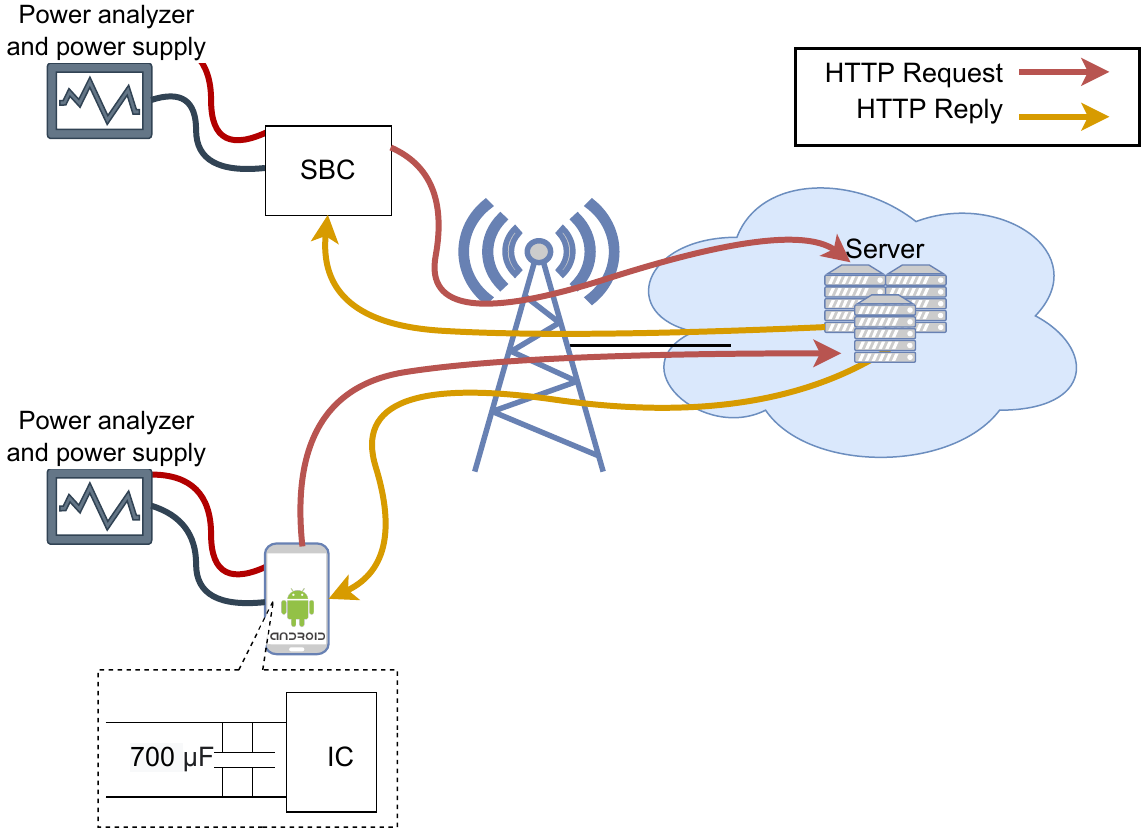}
    \caption{The setup used during the experimental phase.}
    \label{fig:experimetalSetup}
\end{figure}

An overview of the experimental setup is depicted in Figure~\ref{fig:experimetalSetup}, whereas the characteristics of the two devices used in our experiments are shown in Table \ref{tab:devices}.  On the smartphone, the client has been implemented as an Android app relying on OkHttp and cronet.  On the SBC, we used the curl command to issue the HTTP requests~\cite{curl}. We used different versions of curl for sending and receiving HTTP messages. Precisely, we used the standard implementation of curl for message exchanges based on the HTTP/1.1 protocol, nghttp2~\cite{nghttp2} for communications based on the HTTP/2 protocol, and ngtcp2 \cite{ngtcp}  + nghttp3 v.0.6.0-DEV \cite{nghttp3} for HTTP/3-based communication.

The two devices are connected to the Internet via the LTE access network of an Italian Mobile Network Operator (MNO). The SBC gets its network connectivity from an external LTE module \cite{LTEantenna}.
To measure the power consumption of the client devices, we used a Qoitech Otii Arc power analyzer~\cite{otii}. The Otii Arc provides power to the device under test and, at the same time, logs the absorbed current, the supply voltage, and the overall power consumption. To connect the power analyzer to the smartphone we had to make some non-reversible modifications to the device. First, we separated the battery cell from its control circuit. Then, to compensate for any voltage drop at the input of the smartphone, we soldered 7 ceramic capacitors of 100~$\mu$F each, in parallel, as close as possible to the battery controlling circuit. The need to add the capacitors originates from the relatively abrupt changes in the current absorbed by the smartphone: e.g. when the CPU moves to higher frequencies or when subsystems like the cellular modem enter more power-hungry states. In particular, the capacitors help keep the voltage stable. The capacity has been empirically found, starting from the guidelines provided in \cite{otii-capacitor}. Finally, we connected the battery-controlling circuit to the power analyzer. To connect the SBC to the power analyzer, we used the GPIO connectors. We also connected the SBC to the power analyzer via UART. The UART connection allows the SBC and the Otii to communicate. This proved to be particularly useful in synchronizing the operations of the SBC with the power trace collected by the Otii Arc. Synchronization between the smartphone and the Otii Arc was instead not possible: the USB port of the smartphone could not be used, as we did not want to provide additional, and unmeasured, power to the device. Also, wireless communication was not an option, as it would require turning on the Bluetooth or Wi-Fi transceivers of the smartphone, which would introduce some bias in the evaluation.

Since the power analyzer is able to measure the energy needed by the whole device, on a smartphone it is necessary to consider the impact that third-party applications might have. For this purpose, before the beginning of the experimental phase, we performed a factory reset of the device and we removed all the pre-installed apps. Unfortunately, there are a small number of apps that could not be removed. For such apps, we disabled the possibility of using the network when executed in the background. To rule out the energy consumed by the display, we switched it off right after the start of the experiment, and we removed the part of the trace preceding the display switch-off.

For the server side, we used two commercially available edge/cloud solutions: the Cloudflare Workers platform~\cite{cloudflare-worker}, and the Google Cloud Storage service (GCS)~\cite{google}. The two edge/cloud solutions differ in some aspects, summarized below.
The Cloudflare Workers platform allows serverless functions to be executed in one of the 270 Cloudflare data centers. The facility can not be selected by the customers. Instead, the facility used to reply to the client is automatically chosen by the platform on the basis of a proximity criterion~\cite{clouflare-facility}. To limit the server-side processing time, our serverless function implements a simple HTTP echo service, i.e. it returns the just received payload without any modification. The server replies as soon as possible, as the focus is on communication and not on computation. The payload of the first request is generated randomly, while for the subsequent requests only the first 40 bytes are randomly rewritten. This allows to limit the use of computational resources when large payloads are involved. At the same time, having a different payload for each request inhibits possible caching mechanisms that can be present along the path.

The GCS service allows to retrieve files from buckets hosted on Google servers, via the HTTP protocol. In this case, the client periodically sends an HTTP GET request, downloading a given resource from a target bucket. While in the Workers-based scenario, the communication is symmetric, i.e. the request and the response have the same payload, in this case, the communication is asymmetric, i.e. the request does not contain any payload. To prevent any potential caching mechanism, the client asks for a different resource at each request.

\section{Design of Experiments}
\label{sec:design}

The constraints introduced by the specific client device and by the server platform impose some differences in how the experiments have been carried out and in how the collected data have been analyzed. 

In the Cloudflare Workers experiments, the client periodically issues HTTP POST requests. The period is 20 s. We used the \emph{wrangler} tool to collect server-side logs produced by Cloudflare workers. Wrangler is a command-line tool provided by Cloudflare for monitoring and managing a worker. Via the logs generated by wrangler, it is possible to obtain a record concerning each request handled by a worker. Records contain information such as the addresses of the two end-points, the size of the payload, the headers of the request, and the ID code of the facility that handled the request. From the analysis of the logs, we noticed that HTTP requests were served by different facilities.
At the same time, the smartphone is only roughly synchronized with the power analyzer, as there is no way to make the two devices communicate with each other without biasing the results, as previously mentioned. This means that, for the smartphone experiments, it is not possible to estimate the energy consumption of every single HTTP request-response, but only the energy spent in longer periods. As a consequence, the approach was to run experiments with a duration of one hour. This limitation, combined with the impossibility of selecting the facility, made the data collection phase particularly time-consuming: the only possibility was to run experiments and discard all the data that involved a facility other than Milan (the closest to the client device) or whenever a change to a different facility occurred during the 1-hour period.
Each 1-hour experiment was performed during the day. The experiments were repeated several times at different hours. The rationale is that the increased amount of cross traffic that is present during peak hours can possibly impact the results. In more detail, we identified seven 1-hour intervals during the daytime, so that experiments occurring during the same hour are likely to find similar network conditions. 
Because of these limitations, we have been forced to explore a set of payload sizes that is not particularly wide,  with a payload size $s \in \{$400 KB, 1024 KB, 2048 KB$\}$. 

\begin{figure}[!t]
    \centering
    \includegraphics[width=0.7\textwidth]{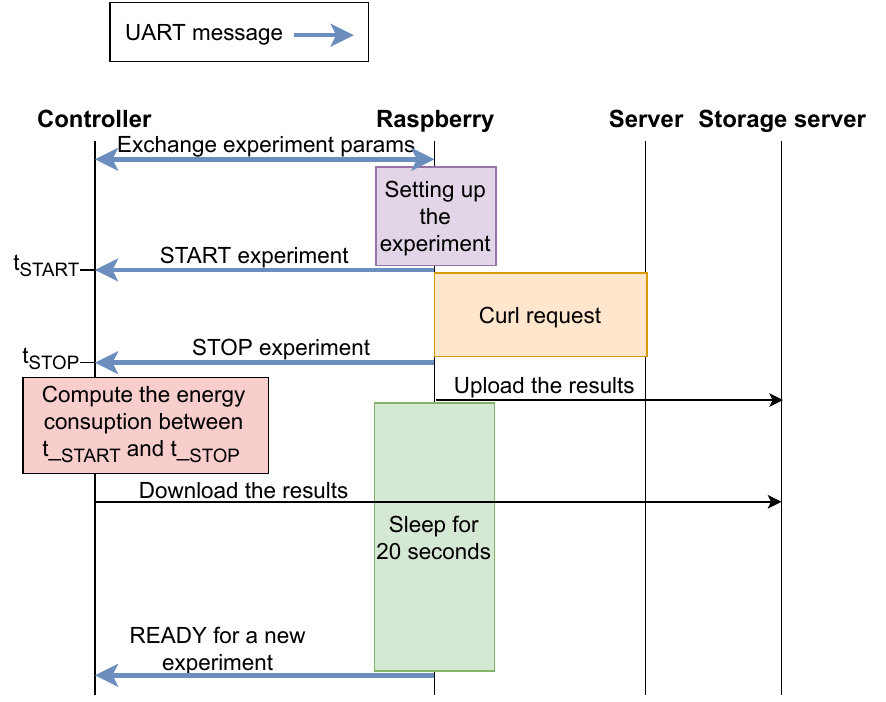}
    \caption{Communication scheme between the controlling software running on the PC connected to the power monitor and the SBC.}
    \label{fig:diagrammaraspi}
\end{figure}

The SBC can communicate with the power analyzer via UART. As a consequence, we were able to estimate the energy spent on every single HTTP request/response. For symmetry reasons, in the SBC-Cloudflare experiments we explored the same set of payload sizes as the smartphone-Cloudflare experiments. However, the analysis was carried out using the energy consumption of the single requests, without the need to average over longer periods as done for the smartphone-based experiments. The communication scheme between the controlling software running on the PC connected to the power monitor and the SBC is depicted in Figure~\ref{fig:diagrammaraspi}.

In the experiments involving the GCS service, we also assessed the impact of latency, by setting up three distinct buckets located in Milan (Italy), South Carolina (USA), and Melbourne (Australia). The Milan facility is relatively close to the client and can be considered as a surrogate of an edge-based solution. The two other facilities are representative of distant, or extremely distant, cloud infrastructure. Since the server can be chosen, experiments could be executed in less time than the Cloudflare ones, thus, for both the smartphone and the SBC, the set of payload sizes that has been explored is larger: $s \in \{ $128 KB, 256 KB, 512 KB, 1024 KB, 2048 KB, 4096 KB$\}$. Since for the SBC-based experiments it is possible to estimate the energy needed by every single request, the three HTTP versions have been executed back to back\footnote{Back to back, but keeping the same 20 s period between a request and the next one.} but in a random order for every given facility and $s$ combination. The controlling software (Controller in Figure~\ref{fig:diagrammaraspi})  sends to the SBC the parameters concerning the experiment to be carried out. The exact timestamps indicating the beginning and end of an HTTP request-response cycle are signaled via UART (the START and STOP messages). Other information logged by the SBC is uploaded to an external storage server (in particular, the verbose output of curl). The Controller downloads information from the external storage and merges the power trace file with the timestamp and additional logged information in a single trace for later analysis. 

\begin{figure*}[!t]
    \centering
    \includegraphics[width=1\textwidth]{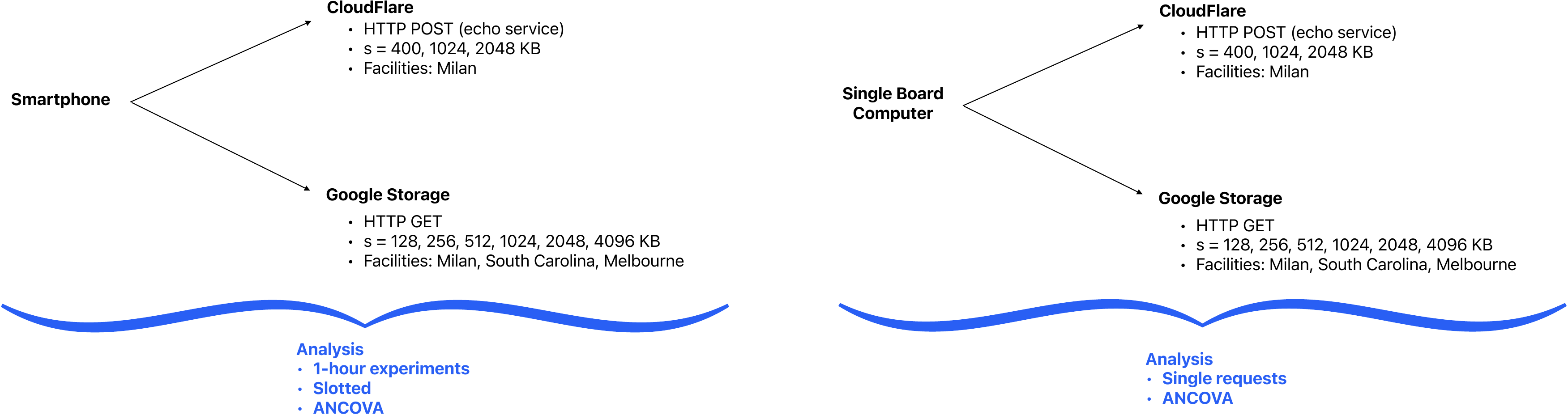}
    \caption{Overview of the experiments.}
    \label{fig:exp-design}
\end{figure*}

Figure \ref{fig:exp-design} summarizes the main characteristics of the experiments, as well as the main analysis criteria for the different client and server combinations.

\section{Results}

We first present the results obtained with the smartphone and then the results obtained with the SBC. 

\subsection{Smartphone-Cloudflare}

\begin{figure}[!t]
    \centering
    \includegraphics[width=0.49\textwidth]{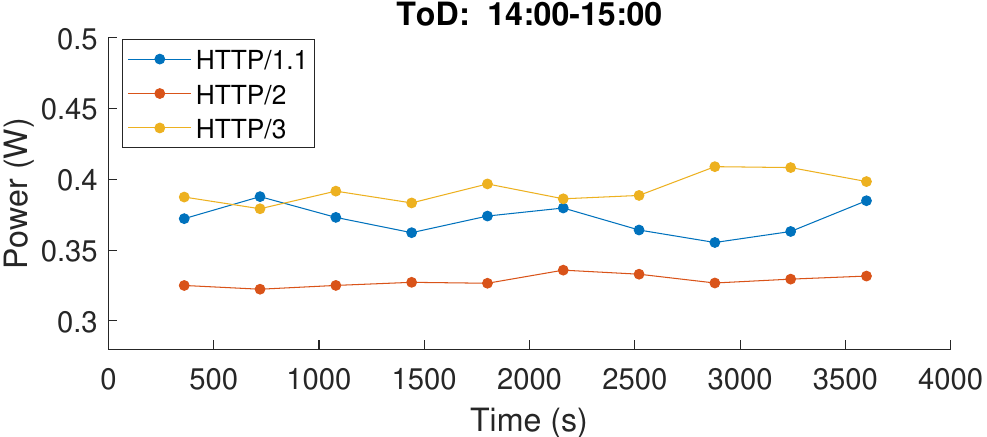}
    
    \caption{The energy consumption of the smartphone when interacting with the serverless edge infrastructure; payload size ($s$) is equal to 1024 kB.}
    \label{fig:Energy_1024}
\end{figure}

For each 1-hour experiment, we processed the metrics collected by the power analyzer as follows. First, we divided the results of the experiments into 10 slots of equal size. Thus, each slot contains 6 minutes of measurements. Next, for each slot, we calculated the average consumption. Figure \ref{fig:Energy_1024} shows a 1-hour experiment, where each curve corresponds to the energy needed by a different version of the HTTP protocol. Points represent the average energy consumption during the considered 6-minute slot. Even though the behavior of the client application does not change during the experiment, some fluctuations can be observed in the amount of needed energy. Such fluctuations are due to changing network conditions and the possibly different load on the edge/cloud infrastructure. Figure \ref{fig:Energy_1024} suggests different consumption levels for the three versions. To statistically assess the differences, we analyzed all the data as follows.

\begin{figure}[!t]
    \centering
    \includegraphics[width=0.47\textwidth]{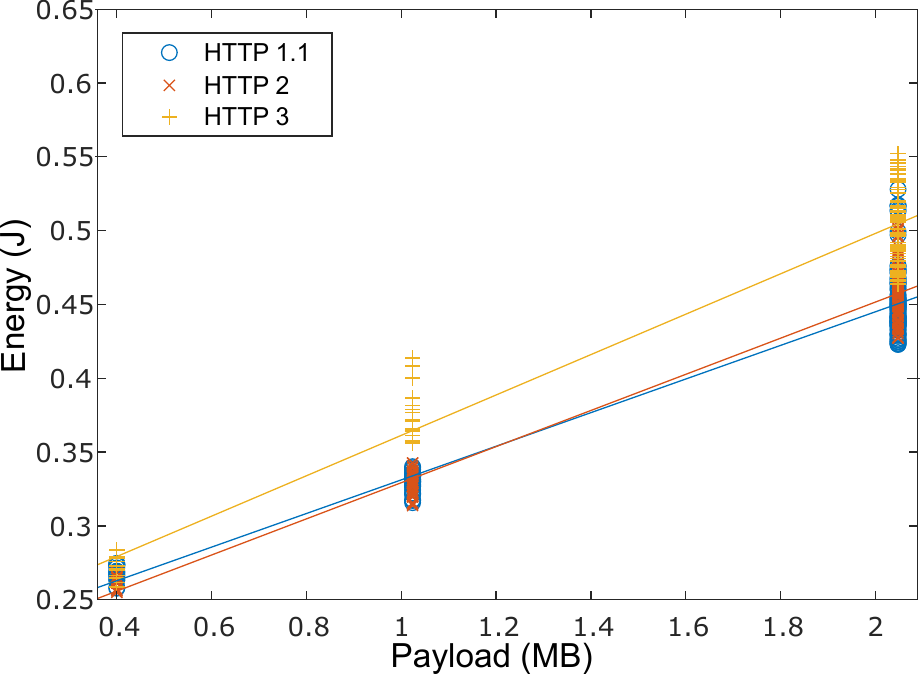}
    \caption{Smartphone-Cloudflare, energy consumption of the different versions of the HTTP protocol, results of ANCOVA analysis. \label{fig:smartphone-cloudflare-ancova}}
\end{figure}

\begin{table}[!t]
\centering
\caption{Smartphone-Cloudflare, average energy consumption (J) for the different HTTP versions. The value between parentheses is the additional energy compared to the most energy-efficient version, in percentage points.}
\begin{tabular}{llll}
\hline
\textbf{Payload (KB)}     & \textbf{HTTP/1.1}   & \textbf{HTTP/2}    & \textbf{HTTP/3}    \\
\hline
400  & 0.263 (+2.7\%) & 0.256 & 0.280 (+9.4\%)\\
1024 & 0.334 (+0.6\%) & 0.332 & 0.365 (+9.9\%)\\
2048 & 0.450  & 0.457 (+1.6\%) & 0.505 (+12.2\%) \\
\hline
\end{tabular}
\label{tab:sm-cl-additional}
\end{table}

We modeled the energy consumption of the three versions of HTTP as a function of the payload size $s$. First, we removed the outliers, defined as the values outside the 10-90 percentile range.
Figure~\ref{fig:smartphone-cloudflare-ancova} shows the results of an analysis of covariance (ANCOVA) where we considered the energy consumption as the dependent variable, \added{the protocol version as the independent variable, and the size of the payload as the covariate. ANCOVA is a statistical technique that is useful to assess if the mean values of the dependent variable (energy) are statistically different across the categories of an independent variable (the three versions of the HTTP protocol) when varying a continuous one (the payload size). The continuous one is called the covariate, as the dependent variable (energy) co-varies with it}. In particular, the energy consumption is modeled as a linear function of $s$ with the coefficients of the linear models possibly varying across protocol versions. The lines corresponding to HTTP/1.1 and HTTP/2 are almost overlapped, with small differences in terms of slopes and intercepts. The line corresponding to HTTP/3 shows a higher consumption compared to the other two versions. We assessed the statistical significance of the difference between the model parameters of HTTP/3 (slope and intercept) compared to the previous versions of the protocol. In particular, we performed multiple pairwise comparisons for all distinct pairs of protocol versions, i.e. HTTP/1.1 vs HTTP/2, HTTP/1.1 vs HTTP/3, and HTTP/2 vs HTTP/3. When comparing a couple of protocols $a$ and $b$, the null and alternative hypotheses are $H_0: m_a = m_b$ and $H_1: m_a \neq m_b$ where $m$ is the slope or the intercept of the line modeling the consumption vs $s$ relationship. Tests have been carried out with $p=0.05$ significance. Results show that the null hypothesis is rejected when comparing the slope of the HTTP/3 model with the other two versions of the protocol. The null hypothesis is also rejected when comparing the intercept of the HTTP/3 and HTTP/2 models. Overall, HTTP/3 needs 9-12\% more energy compared to the previous versions of the protocol (exact values derived from the models are provided in Table \ref{tab:sm-cl-additional}). 

\subsection{Smartphone-GCS}

\begin{figure*}[!t]
    \centering
        \subfloat[Milan (IT)]{
            \includegraphics[width=0.45\textwidth]{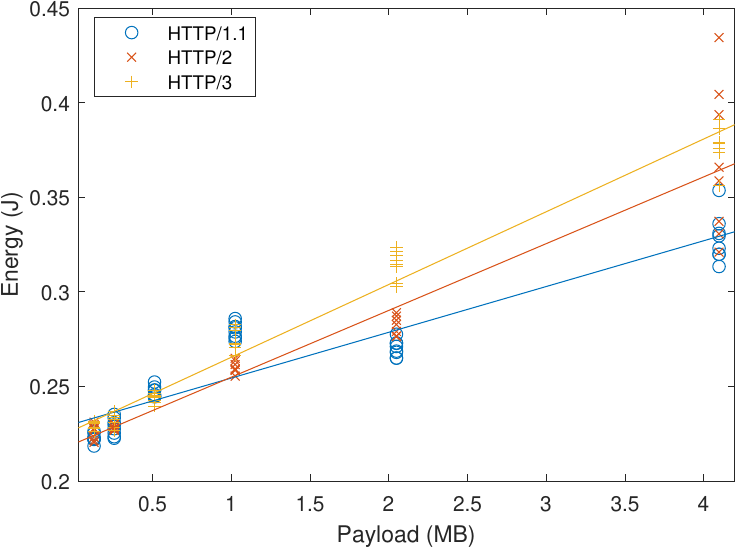}
        }
        \subfloat[South Carolina (USA)]{
            \includegraphics[width=0.45\textwidth]{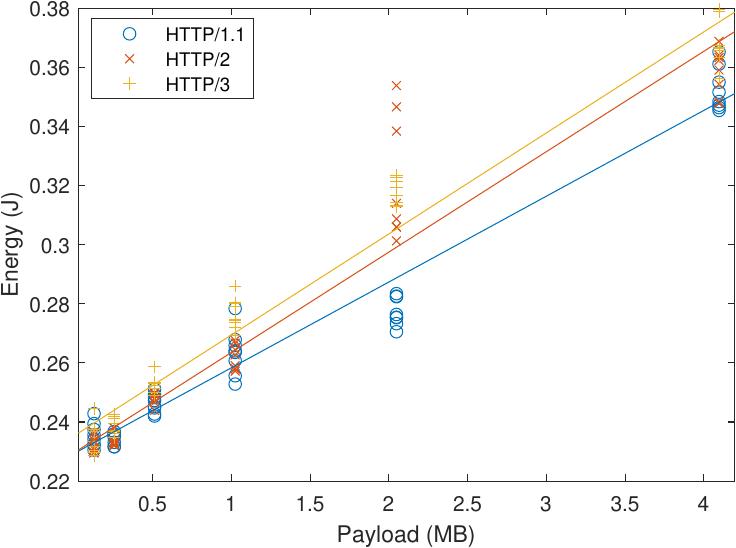}
        }\\
        \subfloat[Melbourne (AU)]{
            \includegraphics[width=0.45\textwidth]{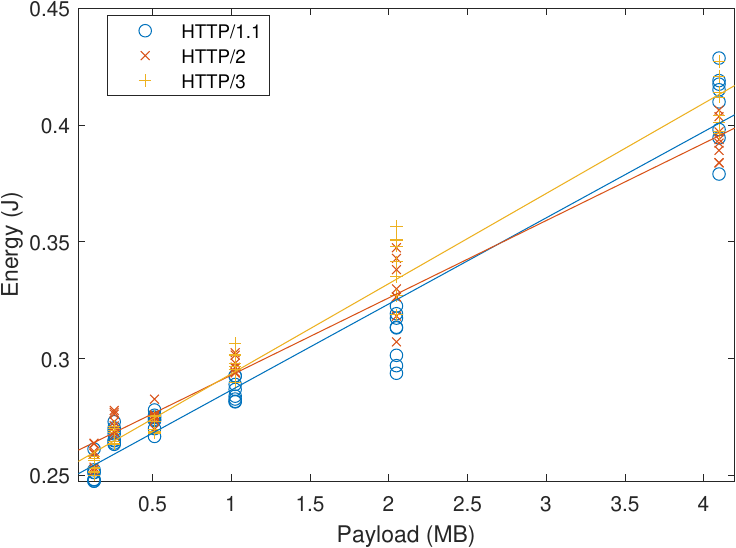}
        }
        \caption{Smartphone-GCS, energy consumption of the different versions of the HTTP protocol, results of ANCOVA analysis.}
        \label{fig:smartphone-GCS}
\end{figure*}

Figure \ref{fig:smartphone-GCS} shows the same ANCOVA-based analysis for the Smartphone-GCS scenario. The analysis has been carried out for the three facilities separately. In general, HTTP/3 requires slightly more energy than the other versions when the facility is not too far (Milan or South Carolina). Instead, when the communication involves a very distant facility (Melbourne), the differences between the three versions tend to disappear. 
In this scenario, the results of the statistical tests do not provide a clear picture: the slope of HTTP/1.1 is different from HTTP/2 and HTTP/3 in both the IT and USA subscenarios. In the AU-based subscenario the null hypothesis is rejected for the slope, with HTTP/2 being less energy-hungry than the other two versions. In the same subscenario, the HTTP/1.1 intercept is lower than the HTTP/2 one according to the significance level. Overall, there there is no clear trend across all the three subscenarios. We also analyzed the whole data, independently from the involved facility, thus ignoring the impact of the physical distance. The average raw results (i.e. the energy values before modeling) are provided in Table \ref{tab:sm-gcs-additional}.

\begin{table}[!t]
\centering
\caption{Smartphone-GCS, average energy consumption (J) for the different HTTP versions. The value between parentheses is the additional energy compared to the most energy-efficient version, in percentage points.}
\begin{tabular}{llll}
\hline
\textbf{Payload (KB)}     & \textbf{HTTP/1.1}   & \textbf{HTTP/2}    & \textbf{HTTP/3}    \\
\hline
 128 & 0.236 & 0.239 (+1\%)& 0.239 (+1\%)\\ 
 256 & 0.243 (+1\%)& 0.242 & 0.246 (+2\%)\\ 
 512 & 0.256 & 0.262 (+2\%)& 0.256\\ 
 1024 & 0.277 (+1\%)& 0.273 & 0.283 (+4\%)\\ 
 2048 & 0.286 & 0.311 (+9\%)& 0.325 (+13\%)\\ 
 4096 & 0.363 & 0.374 (+3\%)& 0.385 (+6\%)\\ 

\hline
\end{tabular}
\label{tab:sm-gcs-additional}
\end{table}

\subsection{SBC-Cloudflare}

\begin{figure}[!t]
    \centering
            \includegraphics[width=0.50\textwidth]{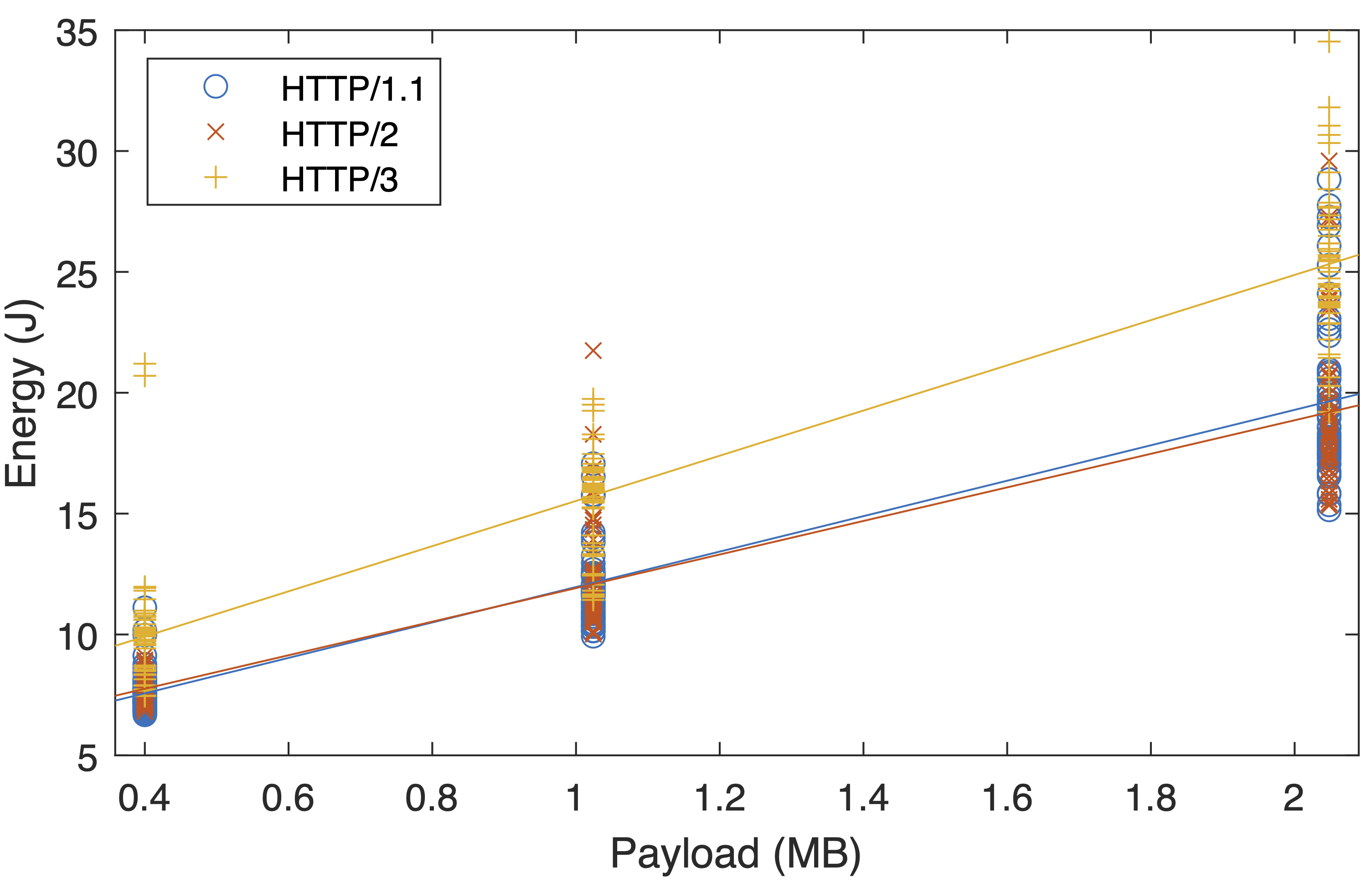}
        \caption{SBC-Cloudflare, energy consumption of the different versions of the HTTP protocol, results of ANCOVA analysis.}
        \label{fig:SBC-Cloudflare}
\end{figure}

\begin{table}[!t]
\centering
\caption{SBC-Cloudflare, average energy consumption (J) for the different HTTP versions. The value between parentheses is the additional energy compared to the most energy-efficient version, in percentage points.}
\begin{tabular}{llll}
\hline
\textbf{Payload (KB)}   & \textbf{HTTP/1.1} & \textbf{HTTP/2} & \textbf{HTTP/3}   \\
\hline
400  & ~7.56          & ~7.75 (+2.5\%)       & ~9.91   (+31.0\%) \\
1024 & 12.13 (+0.4\%)        & 12.09        & 15.75  (+30.3\%)  \\
2048 & 19.64 (+2.3\%)        & 19.20        & 25.32  (+31.9\%)  \\  
\hline
\end{tabular}
\label{tab:sbc-cl}
\end{table}

Figure~\ref{fig:SBC-Cloudflare} shows the results of the ANCOVA analysis carried out on the SBC-Cloudflare scenario. Similarly to the Smartphone-Cloudflare scenario, the energy consumption of HTTP/3 is distinctly higher than in versions 1.1 and 2. Both the slope and the intercept of the HTTP/3 model are significantly different from the ones of the two other versions. Overall, HTTP/3 needs 30\% more energy than the other versions (Table \ref{tab:sbc-cl}). 

\subsection{SBC-GCS}

\begin{figure*}[!t]
    \centering
        \subfloat[Milan (IT)]{
            \includegraphics[width=0.45\textwidth]{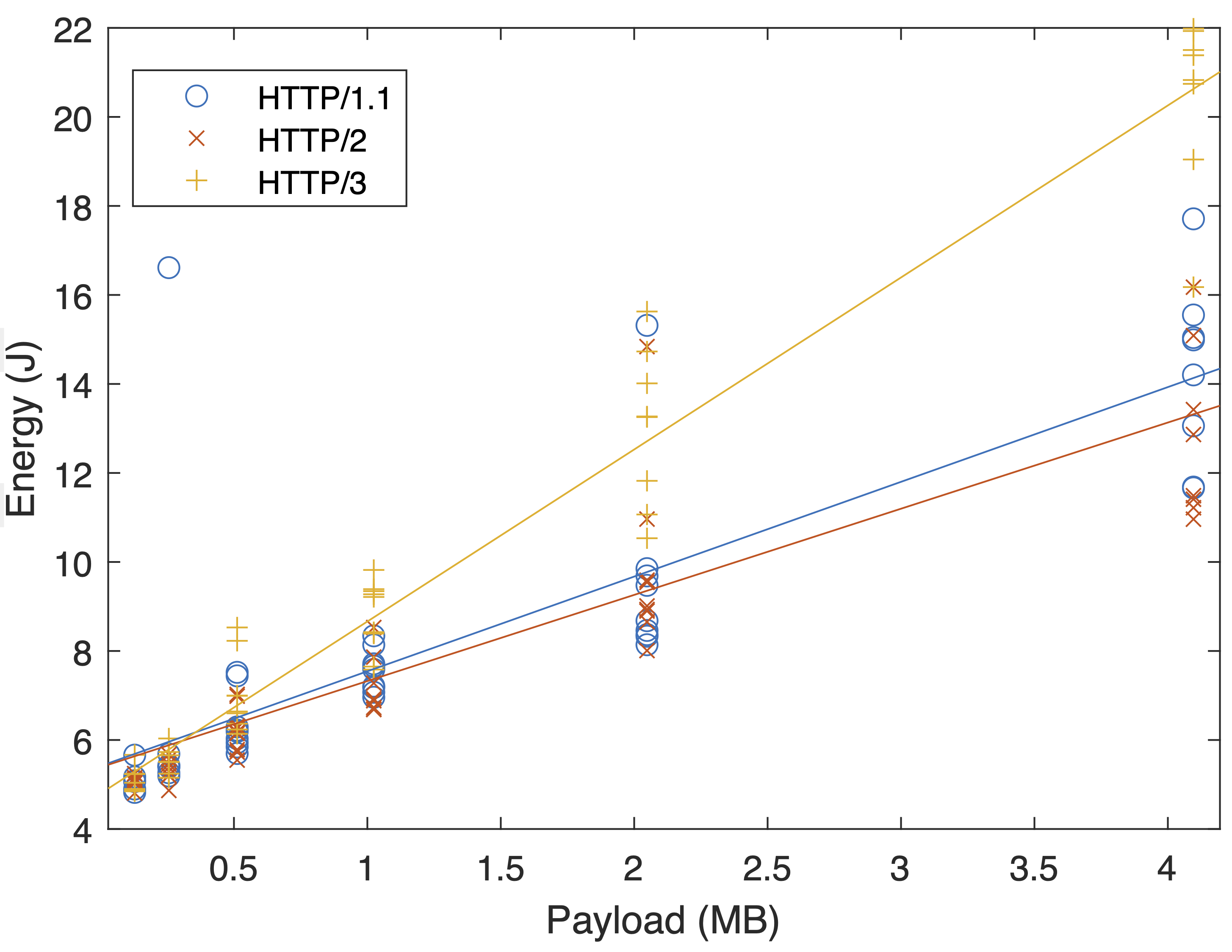}
        }
        \subfloat[South Carolina (USA)]{
            \includegraphics[width=0.45\textwidth]{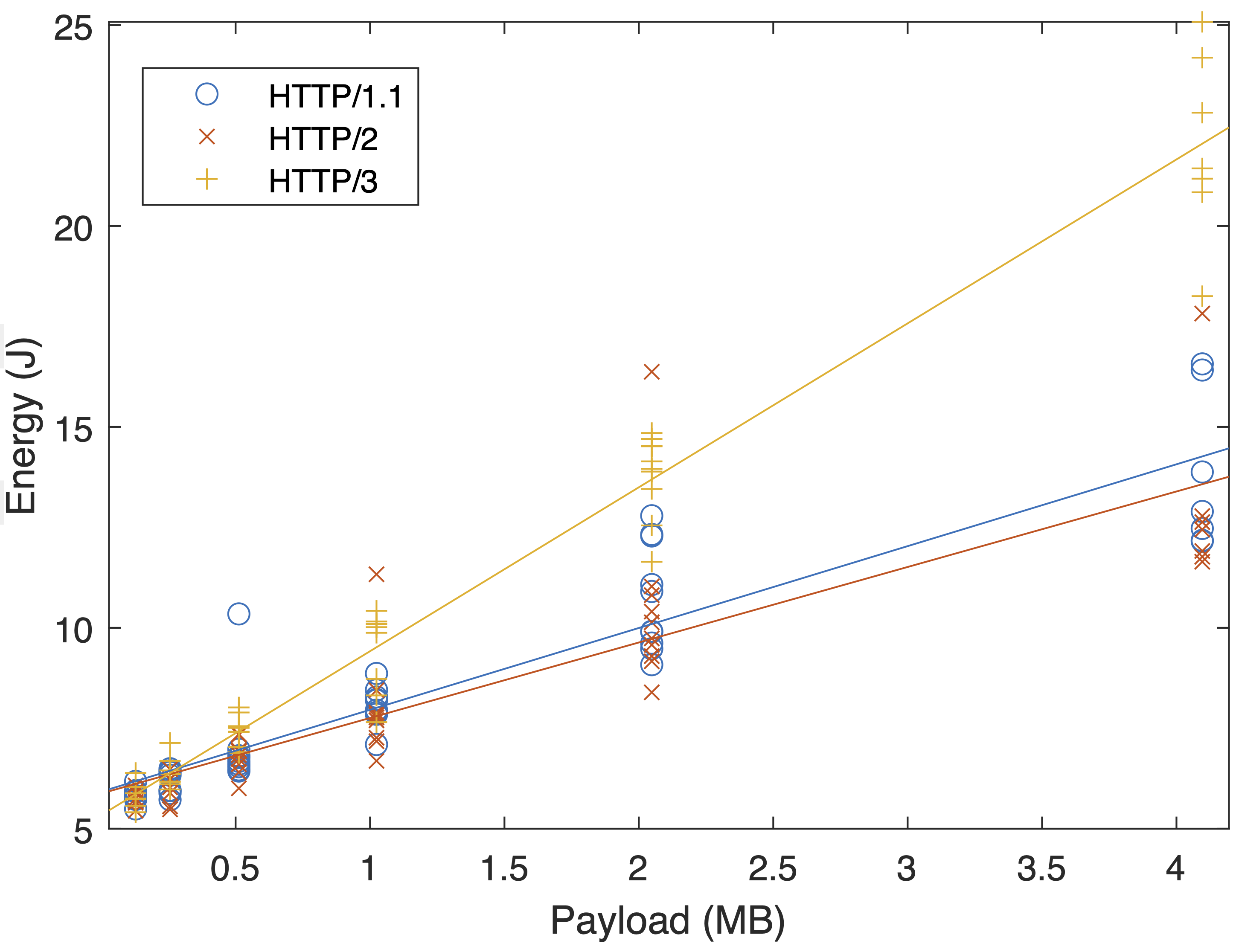}
        }\\
        \subfloat[Melbourne (AU)]{
            \includegraphics[width=0.45\textwidth]{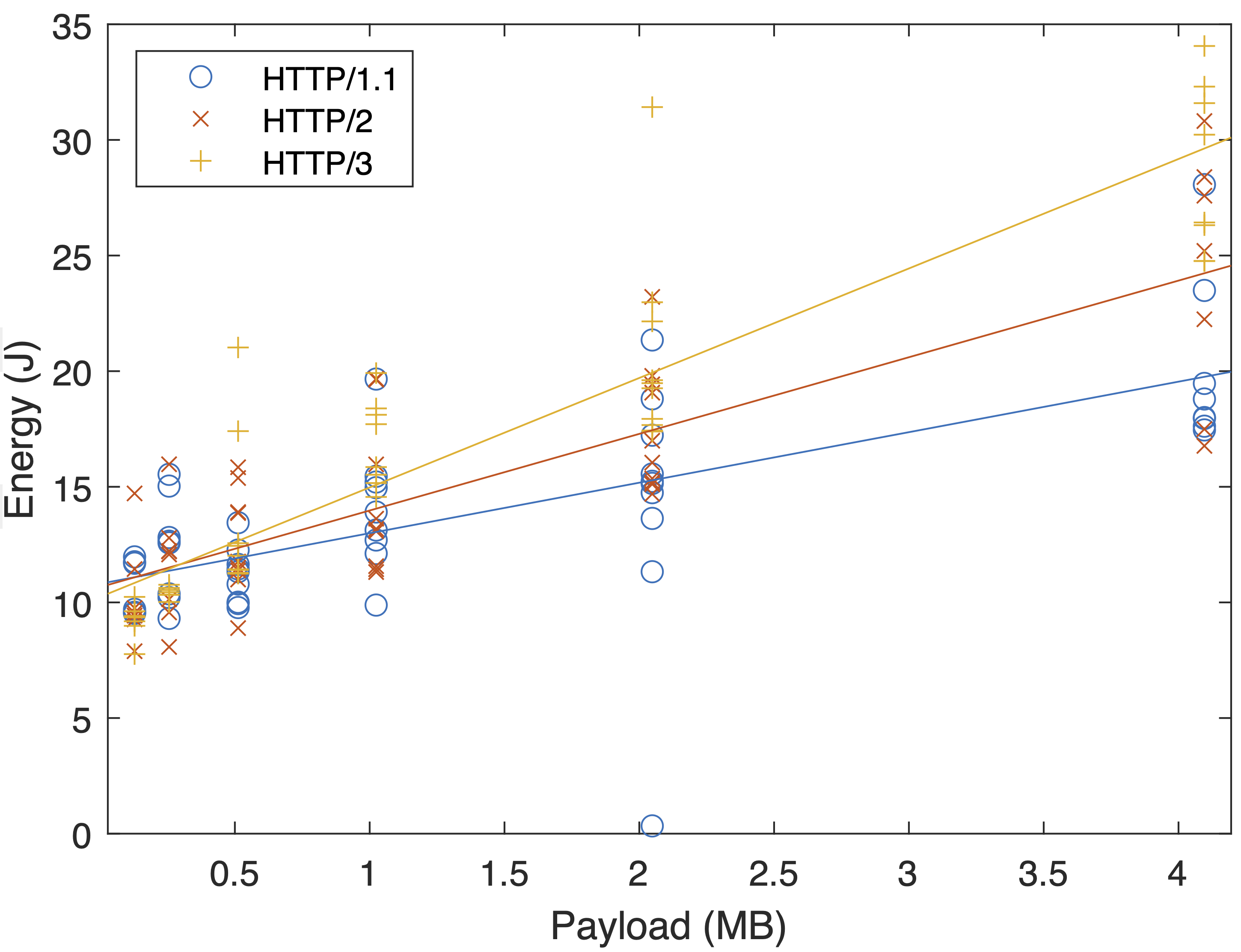}
        }
        \caption{SBC-GCS, energy consumption of the different versions of the HTTP protocol, results of ANCOVA analysis.}
        \label{fig:SBC-GCS}
\end{figure*}

\begin{table*}[!t]
\centering
\caption{SBC-GCS, average$\pm$st.dev. of the energy consumption (J) for the different HTTP versions. The value between parentheses is the additional energy compared to the most energy-efficient version, in percentage points.}
\begin{tabular}{llll}
\hline
\textbf{Payload (KB)}   & \textbf{HTTP/1.1} & \textbf{HTTP/2} & \textbf{HTTP/3}   \\
\hline
 128 & 7.162$\pm$ 2.558 (+4.7\%) & 7.045$\pm$ 2.666 (+3.0\%)& 6.841$\pm$ 1.936\\ 
 256 & 8.419$\pm$ 3.786 (+12.9\%) & 7.659$\pm$ 3.180 (+2.7\%)& 7.458$\pm$ 2.208\\ 
 512 & 8.203$\pm$ 2.401 & 8.542$\pm$ 3.224 (+4.1\%)& 9.265$\pm$ 3.596 (+12.9\%)\\ 
 1024 & 9.914$\pm$ 3.411 & 10.274$\pm$ 4.127 (+3.6\%)& 11.526$\pm$ 3.869 (+16.3\%)\\ 
 2048 & 11.622$\pm$ 4.059 & 12.699$\pm$ 4.226 (+9.3\%)& 15.995$\pm$ 4.388 (+37.6\%)\\ 
 4096 & 16.144$\pm$ 3.957 & 16.461$\pm$ 6.277 (+2.0\%)& 23.819$\pm$ 4.603 (+47.5\%)\\ 
\hline
\end{tabular}
\label{tab:SBC-GCS}
\end{table*}

Figure~\ref{fig:SBC-GCS} shows the results for the SBC-GCS scenario. Again, we carried out a separate analysis for every single facility involved in the experiments. Overall, HTTP/3 again needs more energy compared to versions 1.1 and 2. For example, for a payload of 2 MB, HTTP/3 needs 30\%, 36\%, and 31\% more energy than the HTTP/1.1  protocol when the server is placed in Milan, South Carolina, and Melbourne, respectively. Instead, for a payload of 4MB, HTTP/3 needs 45\%, 53\%, and 50\% more energy than HTTP/1.1. For the more distant facility - Melbourne - all three versions need more energy compared to when shorter paths are involved. 

Table~\ref{tab:SBC-GCS} shows the aggregated raw results across all the facilities (i.e. the energy values before modeling). In this case, also the standard deviation is reported to provide an indication of the aggregated raw results. \added{As can be noticed, for the smallest payloads HTTP/3 performs better than the other versions.} The ANCOVA analysis on the aggregated data highlights that the slope of HTTP/3 is different from the two other protocols with the considered significance level ($p=0.05$).

\subsection{Discussion and Limitations}

\begin{figure*}[!t]
    \centering
        \subfloat[Milan (IT).]{
            \includegraphics[width=0.45\textwidth]{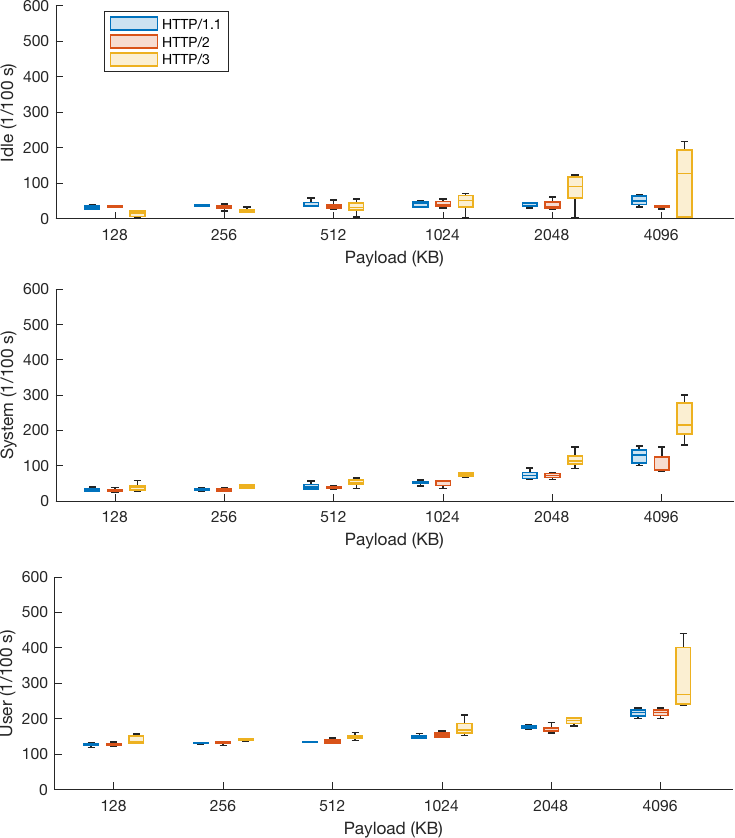}
        }
        \subfloat[South Carolina (USA).]{
            \includegraphics[width=0.45\textwidth]{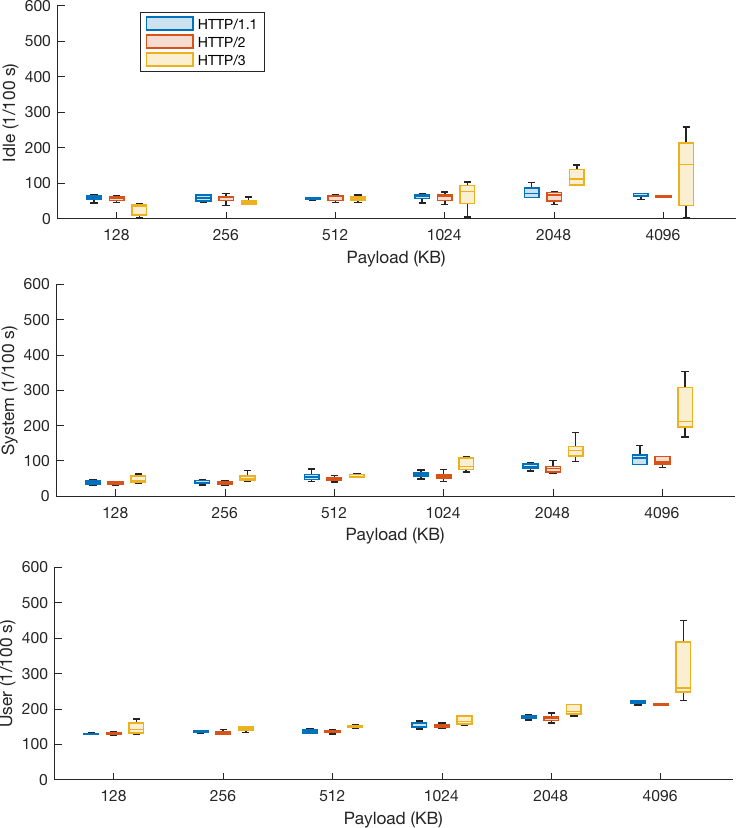}
        }\\
        \subfloat[Melbourne (AU).]{
            \includegraphics[width=0.45\textwidth]{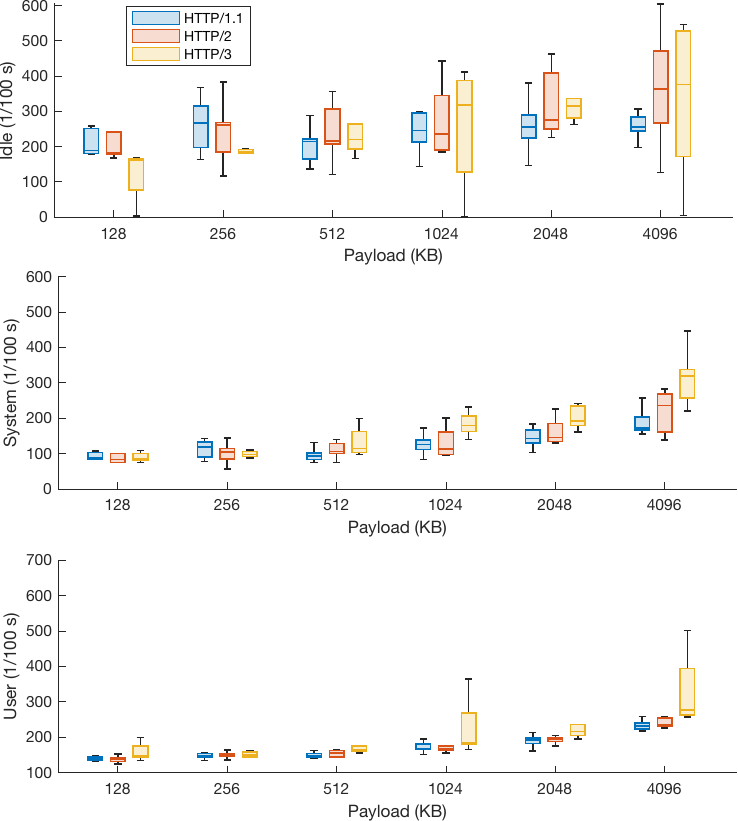}
        }
        \caption{SBC-GCS, CPU utilization for the different HTTP versions and payload sizes.}
        \label{fig:cputime}
\end{figure*}

\removed{Our study focuses}\added{These results are obtained focusing} on machine-to-machine communication and thus\removed{ results} are representative of such a communication style. The experiments assume that the client requests a single resource and it receives the reply, i.e. there is no multiplexing. In addition, requests are sent according to a fixed period. Such a regular pattern is characteristic of automated communication \cite{6629847}. \removed{Further evaluation is needed when multiplexing of requests in is place and in the presence of irregular timing, e.g. when requests are the consequence of Web browsing. We preferred not to focus on interactive-driven communication because when a smartphone is actively used by its owner, the screen is generally the largest energy drain, thus making communication relatively less relevant.}

Given the constraints of the devices in measuring the energy consumption, which imposed the experiments' design choices explained in Section~\ref{sec:design}, isolating the reasons behind our results is challenging. \added{As pointed out in the literature, the parameters that can impact the performance of transport- and application-layer protocols are numerous and not always under our control. Examples range from the adopted congestion control algorithm~\cite{Haile2022:performance, Moulay2021:experimental} and its implementation~\cite{Mishra2022:understanding} to the data transfer size~\cite{Moulay2021:experimental,Shreedhar2022:evaluating}. Even the specific implementation of the protocols can impact~\cite{Dubec2023:performance}.} However, we collected additional measurements to shed light on this aspect. In the following, we present just the results of these measurements for the SBC-GCS scenario.

\begin{figure*}[!t]
    \centering
        \subfloat[Download speed.]{
            \includegraphics[width=0.45\textwidth]{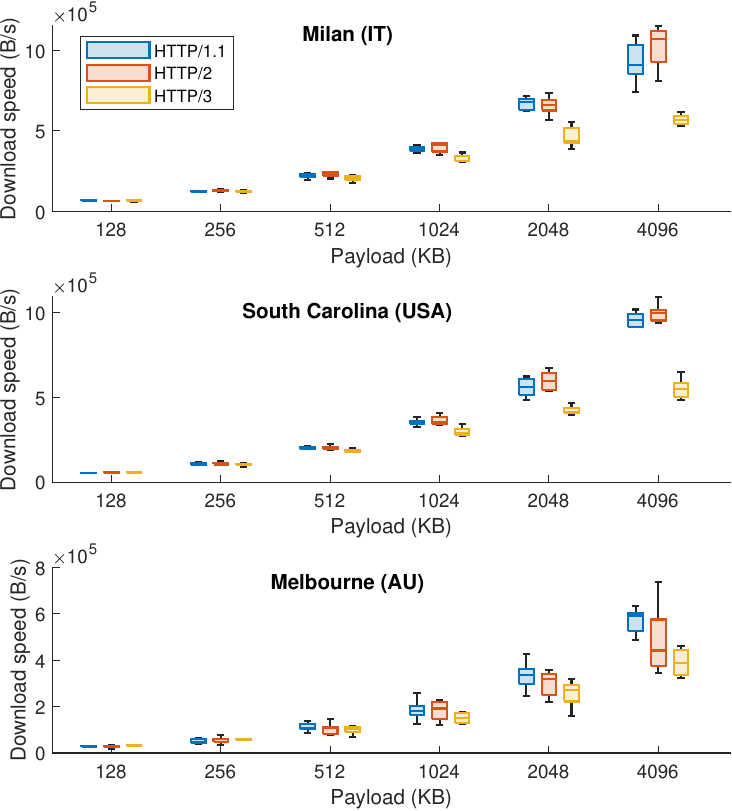}
        }
        \subfloat[Total time.]{
            \includegraphics[width=0.45\textwidth]{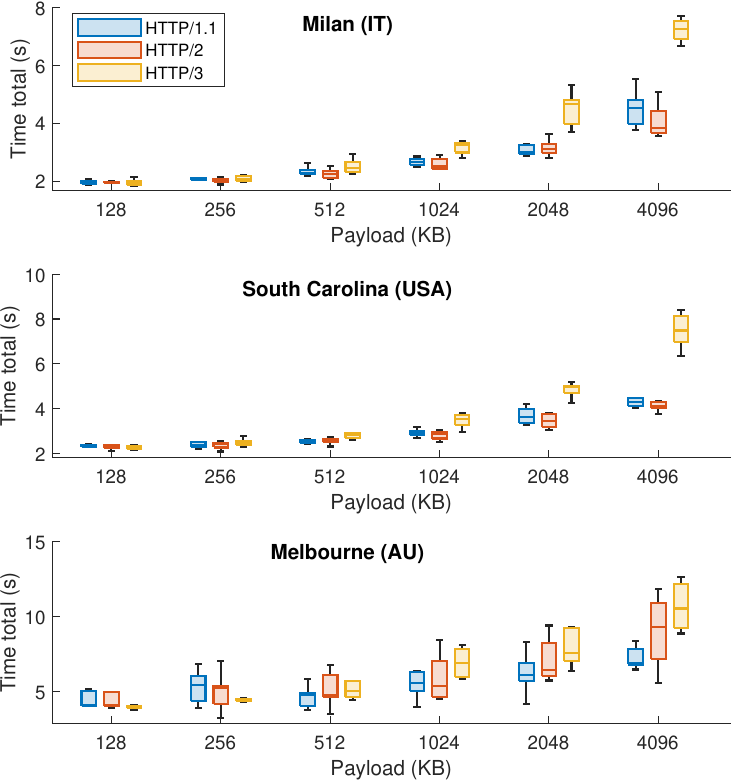}
        }
        \\
        \subfloat[Pretransfer time.]{
            \includegraphics[width=0.45\textwidth]{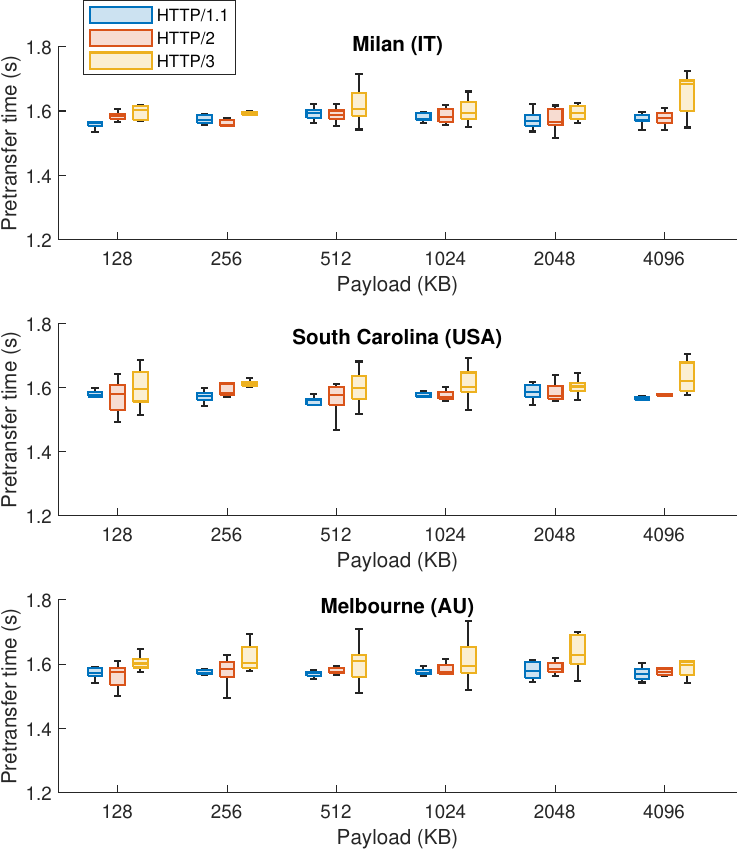}
        }
        \subfloat[Start transfer time.]{
            \includegraphics[width=0.45\textwidth]{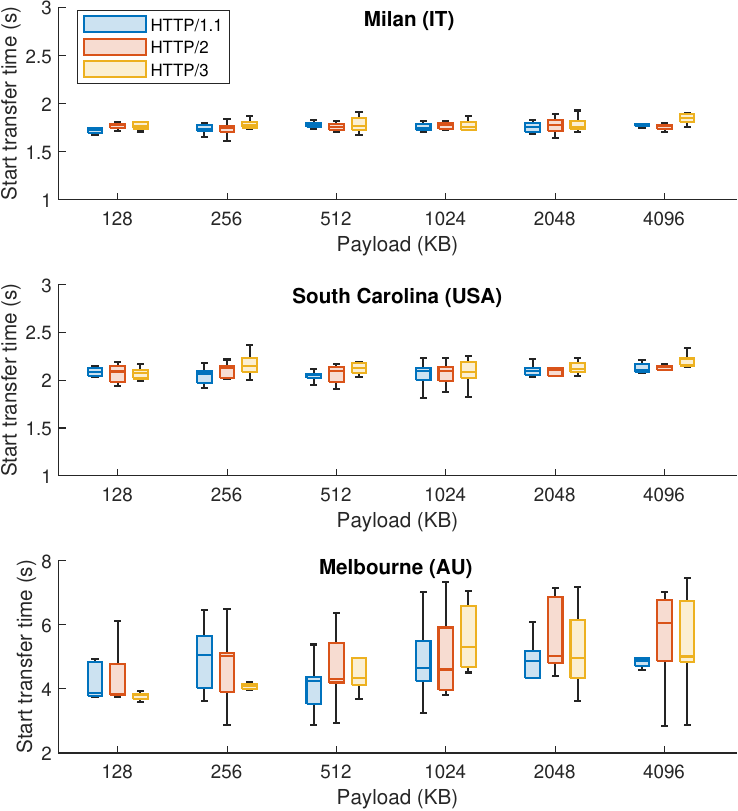}
        }
        \caption{SBC-GCS, download speed, connection time, and total time for each HTTP version and payload size.}
        \label{fig:curlstats}
\end{figure*}

We collected statistics about the CPU utilization time for the different HTTP versions. During the experiments, we parsed the content of the \texttt{/proc/stat} file with a sampling rate of 10 Hz. In particular, we collected the time spent by the CPU in user space, system space, and idle. Figure~\ref{fig:cputime} shows the results, divided per facility and per payload size. We observe that, as expected, the time spent in each state grows with the payload size and with the distance from the facility (\added{since the duration of each request-response increases}). For small payload sizes, HTTP/3 spends less time in the idle state, but as the payload grows HTTP/3 surpasses the other two versions, especially for the two nearest facilities: Milan, and South Carolina. For the third facility, Melbourne, the difference is not so evident, especially if considering HTTP/2 and HTTP/3. This could happen because the influence of the RTT on the idle time is higher for higher RTTs. HTTP/3 spends more time in user space with respect to the other two versions. This is expected, as the underlying transport protocol of HTTP/3 is QUIC, which is implemented in user space. However, also the time spent in system space is higher for HTTP/3, especially for greater payload sizes. \added{These findings help to understand the impact of the different HTTP versions in terms of CPU utilization. More useful insights come from the study published by Shreedhar et al.~\cite{Shreedhar2022:evaluating}, in which the authors discovered that the CPU utilization of QUIC is almost two times the CPU utilization of TCP, reaching up to 90\% utilization. This happens because QUIC is implemented in user space, and it has to deal with packet fragmentation by itself. To avoid fragmentation, QUIC sends UDP packets of limited size. To do so, QUIC spends a significant amount of CPU time in running send and receive message system calls. This does not happen for TCP because the Linux kernel can aggregate packets from a single stream before passing them to the upper layers. Nevertheless, it is important to note that the energy needed depends not only on the CPU utilization (both in user and system mode) but also on the power state of the LTE transceiver, which in turn depends on the transmission and reception pattern of packets.} Overall, it seems that in general HTTP/3 transfers require more time, which could lead to higher energy consumption. To further analyze this aspect, we collected statistics about the data transfer itself from the curl logs. In particular, we extracted the download speed, the total time of the transfer, the pretransfer time, i.e. the time from the start of the operations until the GET request was about to start, and the start transfer time¸ which is the time from the start of the operations until the first byte of the response is received\footnote{Often, in literature, this is referred to as Time To First Byte (TTFB), but we preferred to keep the name given by the curl documentation.}. We show these statistics in Figure~\ref{fig:curlstats}. As can be observed, for HTTP/3 the transfer speed is generally lower than for the other HTTP versions, while the total transfer time is higher, especially for greater payload sizes. \added{This observation is confirmed also by previous literature~\cite{Shreedhar2022:evaluating, Moulay2021:experimental, Dubec2023:performance}, which showed that for big transfer sizes the throughput performance of TCP and older HTTP versions can be higher.} This further suggests that the higher amount of energy consumption is due to higher data transfer times, which could force the LTE interface into more energy-expensive states for longer times. This does not seem to depend on the connection establishment phases, as the pretransfer times of the three HTTP versions are similar. As can be noted, the start transfer time is dependent on the facility, as it involves the transfer of the GET request and the reception of the first byte of the response, which both are influenced by the RTT. In addition, there seems to be a dependence on the payload size, which is more pronounced when the RTT is higher (e.g. in the Melbourne case). However, overall, pretransfer times and start transfer times do not show significant differences among different HTTP versions \added{(it must be noted that these times can be influenced by the specific library implementation, as highlighted in~\cite{Dubec2023:performance})}.

\begin{figure}[!t]
    \centering
            \includegraphics[width=0.6\textwidth]{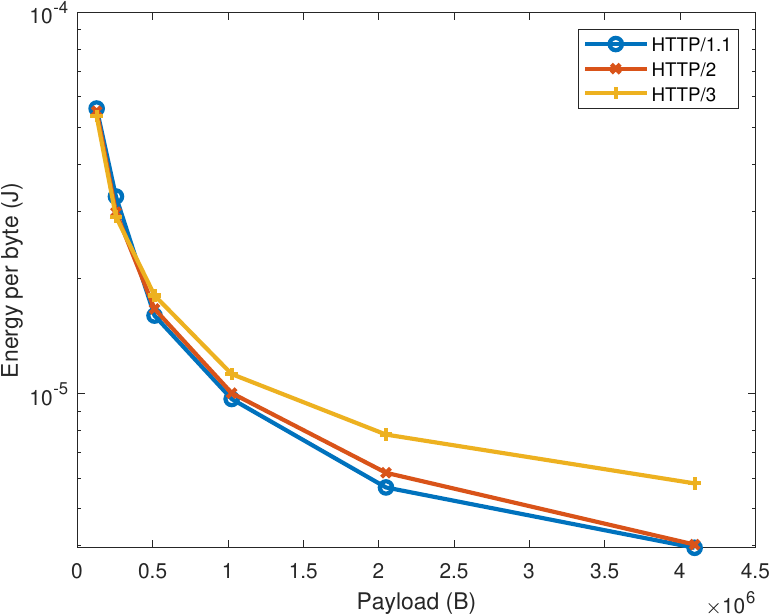}
        \caption{SBC-GCS, energy cost per byte of the three HTTP versions; \added{the y-axis is in log scale}.}
        \label{fig:energyperbyte}
\end{figure}

These findings suggest a last interesting question: what is the energetic cost per unit of information of an HTTP request? We thus considered all the SBC-GCS facilities together, and, for each payload size, we computed the energy consumption per byte for the three HTTP versions. Results are depicted in Figure~\ref{fig:energyperbyte}. It must be noted that these values are not measured but are the result of averaging: we divided the consumed energy of each request-response cycle by the payload size (\added{since the payload size represents the amount of useful information}). First, we can observe that the energy cost per byte decreases as the payload size increases for all the HTTP versions. This is expected, as, for smaller payload sizes, the impact of the overhead (the GET request and the request/response headers) is higher. Thus, from an energy-efficiency perspective, \added{aggregating data into} bigger payloads contributes to saving energy. Most importantly, we can also observe that the energy cost per byte of HTTP/3 is higher than the other two versions, and this is more evident as the payload size increases. It must also be noted that even if the energy per byte does not seem to be extremely high, in modern communication transfers may account for a large amount of data. In this scenario, even a small saving of $10^{-4}$--$10^{-5}$ J per byte can lead to great savings when the accumulated data transfer is in the order of gigabytes.

\section{Browsing-like Evaluation}

As previously mentioned, the focus of this paper is on machine-to-machine communication based on HTTP. However, we also evaluated the energy consumption of the three HTTP versions in a browsing-like scenario. The adopted setup is based on the same SBC used for the experiments described in Section \ref{sec:setup}, and the communication with the controlling PC still takes place according to the scheme shown in Figure~\ref{fig:diagrammaraspi}. We decided to use the SBC for this evaluation, and not the smartphone, because the energy measurements can be more accurate thanks to the UART signaling between the device under test and the smart power supply, in particular, we were able to accurately collect the start and end time of transfers and compute the energy of the corresponding segment in the power trace.  The server side was changed to a Caddy instance running on the University of Pisa premises~\cite{caddy}. The Caddy server was configured to serve static Web pages using all three versions of HTTP. We selected Caddy as it is one of the HTTP/3-ready servers currently available. In addition, it is based on a single executable file, thus reducing the dependency of the server itself on external libraries, which could play a role in the performance of the overall system. The server was executed on an Ubuntu 20.04.6 LTS machine provided with 4 GB of RAM. The static Web pages were stored in the Caddy document root directory. Each page has a size of 2019 kB, which is the median weight of a Web page according to the Web Almanac HTTP Archive’s annual state of the Web report~\cite{httparchive}. Each page included 50 KB of HTML content and ten resources (images) that had to be transferred to the client side. On the client side, the browser-like behavior was implemented through a shell script and the same curl version used in the previously described experiments. In particular, the first execution of curl downloads the main HTML file, the HTML file is then parsed to find all the embedded resources, and finally, all the resources are downloaded. The download of resources is carried out using $N$ parallel HTTP requests. We considered the following values for $N$: 1, 2, and 5. With this approach, we were able to set exactly the number of parallel requests to be used when downloading the embedded resources. We collected the time needed to download the HTML page plus the resources, and the energy required.  Similarly to previous experiments, we evaluated the impact of the distance between the client and the server by adding 0, 50, 100 ms of additional latency for each direction using \texttt{tc}. Caddy was executed with an increased buffer size (416 KB), as it is known that QUIC, and thus HTTP/3 is influenced by such a parameter. 
To avoid possible caching of resources, we set up replicated pages of the same size but with different content. This has been done for every protocol version. Thus, each Web page test was carried out using a different page at every execution of a specific version of HTTP, and using different sets of pages for the three HTTP versions.
We collected at least 25 samples for every combination of $N$, additional delay, and HTTP version. For a given combination of $N$ and additional delay, the three HTTP versions have been used back to back, so that the same configuration is tested for all three versions approximately at the same time. Every time, the three HTTP versions have been executed in a random order to avoid possible distortions introduced by previous executions.
For all the protocols, we removed from the analysis the outliers, defined as the energy values higher than the 90th percentile, to avoid including in the evaluation particularly bad network conditions.

\begin{figure}[!t]
    \centering
            \includegraphics[width=0.6\textwidth]{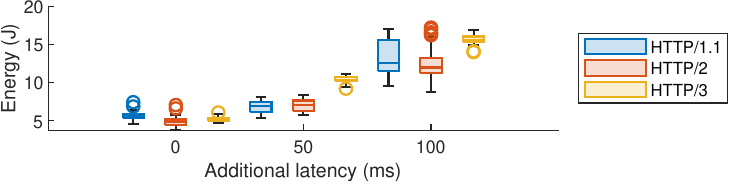}
        \caption{Energy spent to download a 1 MB page containing no resources when varying the latency between the client and the Web server.}
        \label{fig:noresource}
\end{figure}

\begin{figure}[!t]
    \centering
            \includegraphics[width=0.6\textwidth]{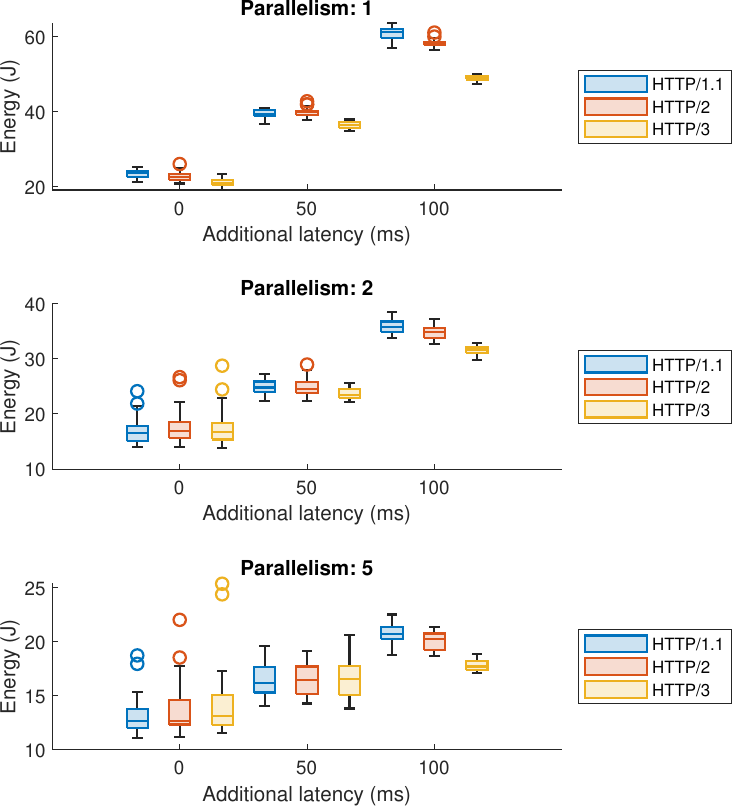}
        \caption{Energy spent to download a page containing ten resources when varying the latency between the client and the Web server; the overall weight of the page is 2019 KB.}
        \label{fig:medium}
\end{figure}

We also executed a preliminary experiment aimed at evaluating the impact of the setup changes compared to the previously presented results. In particular, we run an experiment where the number of resources embedded in the HTML page is zero and the page contains 1 MB of data. The results of this preliminary experiment are shown in Figure~\ref{fig:noresource}. HTTP/3 requires more energy for the high latency sub-scenarios (+50 and +100 ms) whereas its energy consumption is comparable to the other versions when the latency is smaller (+0 ms). Overall, these results are coherent with the ones reported in the previous sections, with HTTP/3 being generally less energy efficient than the other versions of the protocol for that payload size. The main results of the browser-like experiment are shown in Figure~\ref{fig:medium} and the outcome is different from the preliminary experiment. In this case, HTTP/3 proves to be more energy efficient than HTTP/1.1 and HTTP/2, in particular when the parallelism level ($N$) is 1. With increased levels of parallelism (2 and 5), HTTP/3 is still more energy efficient than the other two versions, at least for the high latency sub-scenarios. The major difference between the browsing-like experiment and the preliminary one lies in the execution of multiple requests back-to-back to download the ten resources embedded in the HTML page. Another interesting observation is that increasing the level of parallelism generally reduces the energy needed to download the page and its embedded resources, for all the HTTP versions. This is particularly evident in the high latency sub-scenarios. 

\section{Conclusion}

In this work, we compared the energy consumption of the three main versions of the HTTP protocol. The experimental evaluation was carried out across multiple facilities and using two types of client devices: a smartphone and an SBC. We used a statistical analysis technique to assess the significance of observed differences.

For the machine-to-machine communication pattern, which is the main focus of our study, our results show that over the majority of the experimental scenarios, HTTP/3 is the version that shows the highest energy demands, in particular for the larger payload sizes we considered. Results also show that there is no clear difference between HTTP/2 and HTTP/1.1. As far as we know, this is the first evaluation of the HTTP versions in terms of energy consumption.  Such a general trend is accompanied by significant variations when considering the different types of client devices and service providers. This suggests that implementations and possibly also the interaction between implementations can be responsible for the observed differences across the three versions. The energy-focused analysis was coupled with an evaluation of other performance indicators and computational profiling. Results show that HTTP/3 obtained slower transfer speeds and higher transfer times, which could lead to longer times in more energy-expensive states for the LTE interface. Also, the time spent by the CPU in idle, system, and user depicts a different behavior of the latest iteration of the protocol compared with the previous versions. \added{For some configurations, however, with small payloads HTTP/3 provided slightly better energy efficiency than the other versions. This highlights the importance of selecting the protocol according to the parameters of operation of the specific application and network environment}.

When considering a browsing-like communication pattern, results are quite different. HTTP/3 is almost always the more energy-efficient version of the protocol. This happens as a consequence of back-to-back stacking of requests. In addition, it can be observed that parallelism increases energy efficiency for all HTTP versions.

In this study, we explored a wide variety of scenarios and operational parameters, yet many others could be explored, e.g. configurations parameters of the underlying transport protocols, or the impact of server virtualization/containerization. We believe that these aspects are out of the scope of this study, as the primary goal is to evaluate the client side of energy consumption, and they will be the subject of future studies.

\section*{Acknowledgment}

Work partially supported by the Italian Ministry of Education and Research (MIUR) in the framework of the FoReLab project (Departments of Excellence).

\bibliographystyle{elsarticle-num}
\bibliography{references.bib}

\end{document}